\documentstyle[12pt,a4,epsf]{article}
\makeatletter
\def\fmslash{\@ifnextchar[{\fmsl@sh}{\fmsl@sh[0mu]}}
\def\fmsl@sh[#1]#2{%
  \mathchoice
    {\@fmsl@sh\displaystyle{#1}{#2}}%
    {\@fmsl@sh\textstyle{#1}{#2}}%
    {\@fmsl@sh\scriptstyle{#1}{#2}}%
    {\@fmsl@sh\scriptscriptstyle{#1}{#2}}}
\def\@fmsl@sh#1#2#3{\m@th\ooalign{$\hfil#1\mkern#2/\hfil$\crcr$#1#3$}}
\makeatother
\global\arraycolsep=2pt 
\input{epsf}
\begin{document}
\thispagestyle{empty}
\begin{titlepage}
\begin{flushright}
{\bf TTP95--36} \\
     October 1995  \\
     hep-ph/9510406
\end{flushright}
\vspace{1cm}

\begin{center}
{\Large\bf HADRONIC DECAYS OF } \vspace*{3mm} \\
{\Large\bf EXCITED HEAVY QUARKONIA}
\end{center}
\vspace{0.8cm}

\begin{center}
{\sc Thomas Mannel} and {\sc Res Urech}  \vspace*{2mm} \\
{\sl Institut f\"{u}r Theoretische Teilchenphysik \\
     Universit\"at Karlsruhe \\
     D -- 76128 Karlsruhe, Germany}
\end{center}
\vspace{4cm}
\begin{abstract}
\noindent
We construct an effective Lagrangian for the hadronic decays of a
heavy
excited $s$-wave-spin-one quarkonium into its ground state. We show
that
reasonable fits to the measured invariant mass spectra in the
$J/\psi$ and
$\Upsilon$ systems can be obtained working in the chiral limit. The
mass
dependence of the various terms in the Lagrangian is discussed on
the
basis of a quark model.
\end{abstract}
\end{titlepage}
\newpage
\setcounter{page}{2}
\section{Introduction}
Bound state systems consisting of a heavy quark and a heavy
antiquark
are fairly well described in terms of wave function models, where
the
potential among the two heavy quarks is determined from fitting a
phenomenological ansatz to the data \cite{Buch}.
In this way many static properties
of these systems may be understood even quantitatively.

As far as decays of heavy quarkonia are concerned there has been
also some
theoretical progress recently \cite{BBL,MS95}. For inclusive decays
in
which the two heavy quarks annihilate an effective theory approach
has been developed allowing for a systematic treatment of these
processes.
This method puts the  description of this type of decays on a model
independent basis.

The theoretical description of exclusive hadronic decays is in
general
still quite model dependent. However, for a large class of decay
modes
of excited quarkonia we may use chiral symmetry arguments to
construct
an effective Hamiltonian for decays such as $\Psi' \to \Psi \pi$ or
$\Psi' \to \Psi \pi \pi$ where $\pi$ is a member of the Goldstone
boson
octet. This ansatz has been discussed already in the
literature \cite{Dono}, however, without giving a systematic
derivative
expansion
in the spirit of chiral perturbation theory.

Such an expansion may be performed starting from the infinite mass
limit
for the heavy quarkonium\footnote{%
   The idea we are proposing here is in the same spirit
   as in \cite{Jenkins}, however, in \cite{Jenkins} this
   method is applied to $\Phi$ decays.}.
The momentum of the heavy quarkonium scales with
the heavy quark masses and hence it may not be used as an expansion
parameter. In order to define a systematic derivative expansion
we decompose the momentum $p (p')$ of the heavy
quarkonium in the initial (final) state as
$p (p') = (m_Q + m_{\overline{Q}}) v + k(k')$,
where $k$ and $k'$ are small residual momenta of the order
$\Lambda_{QCD}$,
and $m_Q$ and $m_{\overline{Q}}$ are the masses of the heavy quark
and
antiquark
respectively. The velocity $v$ is the velocity of the initial state
quarkonium and hence the momenta of the final state heavy quarks
will
be of the order of the mass difference of the initial and final
state
quarkonia, which we shall assume to be small compared to the masses
$m_Q$ and $m_{\overline{Q}}$.

In the spirit of chiral perturbation theory one may now formulate a
derivative expansion, which after Fourier transformation becomes an
expansion in the pion momenta and the residual momenta $k$ and
$k'$.
In the next section we shall use chiral symmetry to write down the
leading
as well as the chiral symmetry breaking terms in such an expansion.
In section 3 we show that one may obtain
a good fit to the data, total rates as well as invariant mass
spectra
of the two pions in the decay $\Psi' \to \Psi \pi \pi$. We shall
compare the charmonia and the bottomonia and discuss the dependence
of the
parameters in the effective Lagrangian of the bottom and charm mass
in
section 4. Conclusions are given in section 5.

\section{Effective Lagrangian for Heavy Quarkonia \newline Decays}
We construct an effective Lagrangian for the decays of a heavy
excited $s$-wave-spin-one quarkonium into the ground state, which
is
either a $J/\psi(1S)$ or an $\Upsilon (1S)$. Let $A_\mu$ be the
field of
the excited $s$-wave $1^-$ state and $B_\mu$ the one of
$1^-$ ground state. We shall consider here only quarkonia where
$m_Q = m_{\overline{Q}}$, hence we look at charmonia or bottomonia
systems.

The momenta of the heavy quarkonia are split into a large piece
which
scales as $m_Q$ and a small residual part $k$ which depends only
weakly on
the heavy quark mass
\begin{eqnarray*}
 p &=& 2 m_Q v + k \mbox{ for } B_\mu  \\
 p' &=& 2 m_Q v + k' \mbox{ for } A_\mu
\end{eqnarray*}
In coordinate space this is achieved by a phase redefinition of the
fields
\begin{eqnarray}
A_\mu (x) = \exp (-i2m_Q v \cdot x) A_\mu^{(v)} (x) \nonumber \\
B_\mu (x) = \exp (-i2m_Q v \cdot x) B_\mu^{(v)} (x)
\end{eqnarray}
such that the derivative acting on the fields with superscript
$(v)$
is now $k$ (or $k'$) and small compared to the heavy mass.
The fields $A^{(v)}$ and $B^{(v)}$ obey the equations of motion
corresponding to static fields and are transverse with respect to
the velocity vector $v$
\begin{eqnarray}
iv \cdot \partial A_\mu (x) = 0,  \quad v^\mu A_\mu (x) = 0 \\
iv \cdot \partial B_\mu (x) = 0, \quad v^\mu B_\mu (x) = 0
\end{eqnarray}

We are interested in an effective Lagrangian for the hadronic
decays
of the type $A \to B \pi$ or $ A \to B \pi \pi$, and we construct
this
Lagrangian using chiral symmetry. Higher order terms may be
included by
chiral perturbation theory, i.e.\ by a systematic expansion in the
derivatives
(these correspond to pion momenta or residual momenta for
quarkonia) and
in the light quark masses. The heavy quarkonia are singlets under
the chiral
symmetry, and hence the decays $A \to B \pi$ are forbidden in the
chiral
limit.

In the chiral limit only decays to an even number of pions are
allowed,
and the leading term in the derivative expansion obeying chiral
symmetry
consists of three terms
\begin{eqnarray} \label{l0}
{\cal L}_0&=& g A_\mu^{(v)} B^{(v)\mu *}
             \mbox{ Tr}[(\partial_\nu U) (\partial^\nu U)^\dagger]
           + g_1 A_\mu^{(v)} B^{(v)\mu *}
             \mbox{ Tr}[(v\cdot\partial U) (v\cdot\partial
U)^\dagger]
             \nonumber\\
         &&+ g_2 A_\mu^{(v)} B_\nu^{(v)*}
             \mbox{ Tr}[(\partial^\mu U) (\partial^\nu U)^\dagger +
                        (\partial^\mu U)^\dagger (\partial^\nu U)]
           + \mbox{ h.c.}
\end{eqnarray}
where $U$ is a unitary $3\times 3$ matrix that contains the
Goldstone fields
\begin{equation}
U = \exp\,(i\Phi/F_0)\hspace{1cm}
\Phi= \sqrt{2}\left(
\begin{array}{ccc}
\frac{1}{\sqrt{2}}\pi^0+\frac{1}{\sqrt{6}}\eta_8  & \pi^+ & K^+\\
\pi^- & -\frac{1}{\sqrt{2}}\pi^0+\frac{1}{\sqrt{6}}\eta_8 & K^0\\
K^- & \overline{K^0} & -\frac{2}{\sqrt{6}}\eta_8
\end{array}
      \right)
\end{equation}
and $F_0$ is the pion decay constant\footnote{%
    We use the convention in which the pion decay constant is
    $ F_0 \simeq 93$ MeV.}
in the chiral limit.
In previous considerations the second term in (\ref{l0}) has been
often
omitted, but
since a consistent definition of the derivative expansion forces us
to
introduce the vector $v$, the second term is a perfectly allowed
one.
We shall show in the next section that the magnitude of the
coupling
$g_1$ is not small compared to $g$, at least not for the case of
charmonium, and that a reasonable fit to the invariant mass spectra
of
the pions may be obtained from (\ref{l0}).

The quark mass matrix breaks chiral symmetry, and the leading terms
of this kind are
\begin{eqnarray}\label{lsb}
{\cal L}_{S.B.} &=& g_3 A_\mu^{(v)} B^{(v)\mu *}
                        \mbox{ Tr}[{\cal M} (U +
U^\dagger-2)]\nonumber \\
                && + i g' \epsilon^{\mu \nu \alpha \beta} \left[
                     v_\mu A_\nu^{(v)} \partial_\alpha
B_\beta^{(v)*} -
                     (\partial_\mu A_\nu^{(v)}) v_\alpha
B_\beta^{(v)*}
                     \right]
                     \mbox{ Tr}[{\cal M} (U - U^\dagger)]\nonumber
\\
                && + \mbox{ h.c.}
\end{eqnarray}
 where
\begin{equation}
{\cal M} = \left( \begin{array}{ccc}
                    m_u & 0   & 0 \\
                      0 & m_d & 0 \\
                      0 & 0   & m_s
                     \end{array} \right)
\end{equation}
The two operators proportional to the $\epsilon$ tensor are
relevant for
decays involving an odd number of pions. The relative sign of the
these
terms is fixed by reparameterization invariance \cite{luke92}. An
analogous formulation for the case $V'\rightarrow V\pi$ where
$V',V$ are
members of the lowest-lying vector meson nonet has been given
recently
by Jenkins, Manohar and Wise \cite{Jenkins}.

\section{Fit to the Data}
Starting from the Lagrangian given in (\ref{l0}) we fit the
experimental
data on the decays $\psi(2S) \rightarrow J/\psi\pi^+\pi^-$ and
$\Upsilon(2S)\rightarrow \Upsilon(1S)\pi^+\pi^-$. The amplitude has
the
form
\begin{eqnarray}
{\cal A}(\Psi' \to \Psi \pi^+\pi^-) &=&
-\frac{4}{F_0^2}\left\{\;\left[\,
              \frac{g}{2}(m_{\pi\pi}^2 - 2 M_\pi^2)
            + g_1 E_{\pi^+} E_{\pi^-}\right]
              \epsilon_\Psi^* \cdot \epsilon_{\Psi '} \right.
\nonumber \\
&&\hspace{1.5cm} + g_2 \left[ p_{\pi^+\mu} p_{\pi^-\nu}
                       + p_{\pi^+\nu} p_{\pi^-\mu}\right]
                       {\epsilon_\Psi^*}^\mu {\epsilon_{\Psi
'}}^\nu \biggr\}
\nonumber \\
\end{eqnarray}
where $m_{\pi\pi}= (p_{\pi^+} + p_{\pi^-})$, $p_{\pi^+} =
(E_{\pi^+},{\bf p}_{\pi^+})$ and $\epsilon_\Psi, \, \epsilon_{\Psi
'}$ are
the polarization vectors of the heavy spin-one quarkonia. First we
observe that
the term proportional to $g_2$ describes the $d$-wave part of the
pion
system, which is found to be suppressed experimentally in the
charmonium
decay \cite{abrams75} as well as in the bottomonium decay
\cite{albrecht87} to a few percent. This observation is supported
in
addition by theoretical arguments, see next section. Thus we put
$g_2=0$. The remaining coupling
constants $g,g_1$ can be fitted to the total decay rate and the
invariant
mass spectra of the two pions. The data for the differential rate
are
taken from Fig.7 in
\cite{albrecht87}.
\begin{figure}[p]
 \begin{center}
   \vspace*{-1.5cm}
   \epsfxsize=13.1cm
   \leavevmode
   \epsffile[0 0 624 482]{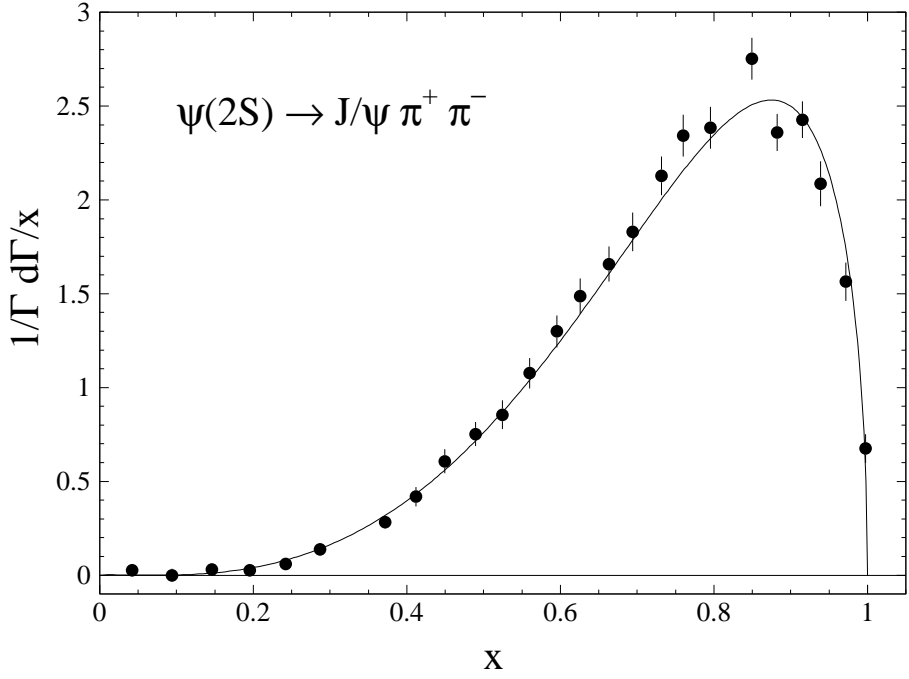}\\
   \epsfxsize=13.1cm
   \leavevmode
   \epsffile[0 0 624 482]{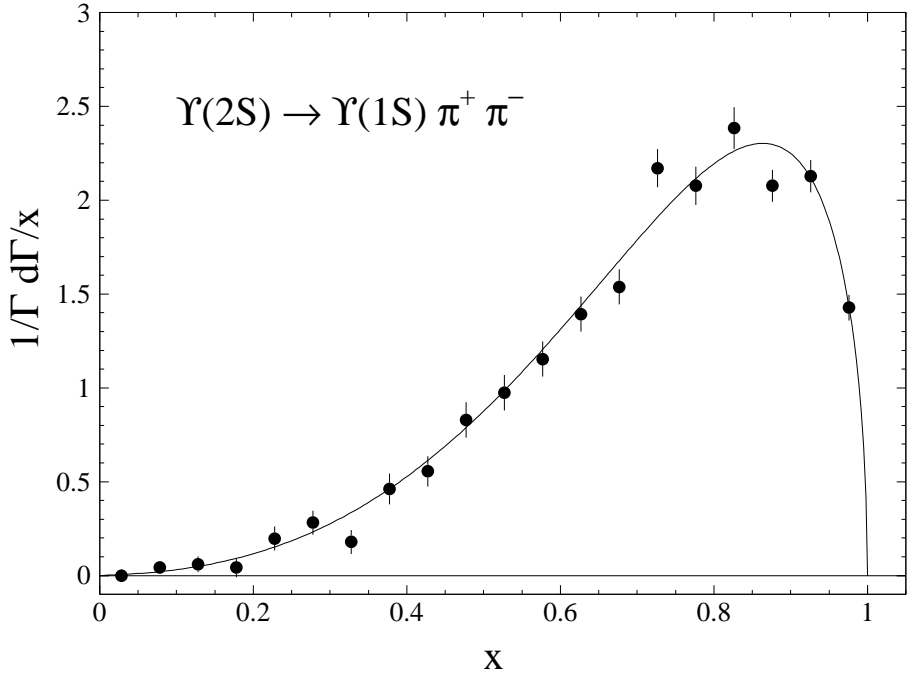}
 \caption[]{The curves show the fit to the invariant mass spectra
of the
two pions with $x=(m_{\pi\pi} - 2M_\pi)/(M_{2S} - M_{1S} - 2M_\pi)$
and
$m_{\pi\pi}=(p_{\pi^+} + p_{\pi^-})$. The
experimental data are taken from \cite{albrecht87}.}
 \end{center}
\end{figure}
For the total decay rate we used the PDG values \cite{pdg94}. Our
results are listed in Table \ref{tab1}, and the fits are shown in
Fig.1.
\begin{table}[t]
$$
\begin{array}{c|c|c|c}
\hspace{2cm}&         &                             &\hspace{3cm}
\\[-2mm]
    & \psi(2S) \rightarrow J/\psi\pi^+\pi^- & \Upsilon(2S)
                 \rightarrow \Upsilon(1S)\pi^+\pi^- & \chi^2/d.f.
\\[2mm]
\hline &&&\\[-3mm]
g   & 0.30 \pm 0.02   & 0.25 \pm 0.02               & 26.2\,/\,24
\\[2mm]
\hline &&&\\[-3mm]
g_1 & -0.11 \pm 0.01  & -0.04 \pm 0.01              & 26.3\,/\,19
\\[2mm]
\end{array}
$$
\caption[]{Values of the coupling constants obtained from a fit to
the data
as given in \cite{albrecht87}. The errors quoted are purely
experimental.}
\label{tab1}
\end{table}
We find that the values of $g$ in both systems are very close to
each
other,
whereas the ratios $g_1/g$ differs by a factor of 2, namely
\begin{equation}
\left(\frac{g_1}{g}\right)_{c\bar{c}}=-0.35 \pm 0.03  ,\hspace{2cm}
\left(\frac{g_1}{g}\right)_{b\bar{b}}=-0.18 \pm 0.02
\end{equation}
With the inclusion of the symmetry breaking part (\ref{lsb}) the
situation
does not change at all. If we consider the mass term as a
perturbation to
the operators relevant in the chiral limit, the best fit is again
obtained
for $g_3 =0$\footnote{In the case of charmonium $g_3$ is slightly
             different from zero, but still beyond the accuracy to
             which theresults are given.}.
The observation, that the data can be fitted
accurately with operators allowed in the chiral limit, has been
made by
Novikov and Shifman \cite{novikov81}. To our knowledge, however,
the
rigorous description in the language of effective Lagrangians has
not
been presented before.

In the case of the decay amplitudes of $A\rightarrow B \pi$ where
$\pi=
\pi^0, \eta$ experimental values exist in the charmonium system,
for
bottomonia upper limits are known only. We consider the
coupling $g'$
in the Lagrangian (\ref{lsb}) with respect to the same decay
channel in the
different systems. The stronger limit is found in the $\eta$ decay
channel
\begin{equation}
\left(\frac{g'_{b\bar{b}}}{g'_{c\bar{c}}}\right)_\eta < 0.6
\end{equation}

\section{Discussion of the Results}
In this section we shall discuss our results, in particular the
mass
dependence of the coefficients $g$, $g_1$ and $g'$. We shall try to
achieve an at least qualitative understanding by representing the
coefficients in the Lagrangian by matrix elements of heavy quark
operators. Hence we consider a model in which the heavy quarks and
the
light pseudoscalar mesons are the relevant degrees of freedom; in
this
respect this resembles the chiral quark model of Georgi and
Manohar \cite{ChiQua}.

First of all, this model easily explains the fact that the
pion-$d$-wave contribution is suppressed, since this at least
requires a
dimension-6-operator for the heavy quarks, which would be contained
in the operator
\begin{equation}
{\cal L}_{d-wave} = \frac{a_2}{\Lambda^3} (\bar{Q}_v \gamma_\mu
Q_v)
                    (\bar{Q}_v \gamma_\nu Q_v)
\mbox{Tr } \left[
           (\partial^\mu U)(\partial^\nu U^\dagger)
          +(\partial^\nu U)(\partial^\mu U^\dagger) \right]
\end{equation}
while the pion-$s$-wave
component may be represented by a dimension-3-operator for the
heavy quarks
\begin{equation}
{\cal L}_{s-wave} = a_0 (\bar{Q}_v Q_v)
\mbox{Tr } \left[
           (\partial^\mu U)(\partial_\mu U^\dagger) \right]
\end{equation}
Here $Q_v$ are the operators of heavy quarks which have been
rescaled with a phase according to
$$
Q_v = \exp (imvx ) Q
$$
and the scale $\Lambda$ is at least of the order of the chiral
symmetry
breaking scale or maybe even the heavy quark mass; in the latter
case the $d$-wave contribution to this decay would be more strongly
suppressed in the $\Upsilon$ decays as compared to the $J/\psi$
transitions.

We have argued that the $d$-wave contribution is suppressed and
hence
we have only two operators left in the chiral limit. Both operators
contribute to the $s$-wave of the decays $\psi(2S) \to J/\psi \pi
\pi$ and
$\Upsilon(2S) \to \Upsilon (1S)\pi \pi$. Employing the model
discussed
above, we have the corresponding two operators on the quark level
\begin{eqnarray}
{\cal L}_{quark} &=& a_0 (\bar{Q}_v Q_v)
\mbox{Tr } \left[
           (\partial^\mu U)(\partial_\mu U^\dagger) \right] \\
\nonumber  &+& b_0 (\bar{Q}_v Q_v)
\mbox{Tr } \left[
           (v \cdot \partial U)(v \cdot \partial U^\dagger) \right]
\end{eqnarray}
The coupling constants $g$ and $g_1$ are related to the couplings
in
the quark Lagrangian ${\cal L}_{quark}$ through the relations
\begin{eqnarray}
&& g \langle \Psi | A^\mu B_\mu^* | \Psi ' \rangle =
   g \epsilon_\Psi^* \cdot \epsilon_{\Psi '}
=  a_0 \langle \Psi | \bar{Q}_v Q_v | \Psi ' \rangle  \\
&& g_1 \langle \Psi | A^\mu B_\mu^* | \Psi ' \rangle =
   g_1 \epsilon_\Psi^* \cdot \epsilon_{\Psi '}
=  b_0 \langle \Psi | \bar{Q}_v Q_v | \Psi ' \rangle
\end{eqnarray}
{}From the combined chiral and
heavy mass expansion one would expect that both coupling constants
$a_0$ and $b_0$ scale in the same way with the heavy quark mass. As
far as the mass dependence of $g$ and $g_1$ is concerned, one has
to
take into account the mass dependence of the matrix element
$\langle \Psi |
\bar{Q}_v Q_v | \Psi ' \rangle$. For the operators itself there
exists
a heavy mass limit, while the mass dependence of the states may not
be
accessed within the framework of the $1/m_Q$ expansion \cite{MS95}.
To this end we have to rely on simple dimensionality arguments.
Assuming that the states are normalized to unity we may write
the matrix elements as
\begin{equation}
\langle \Psi | \bar{Q}_v Q_v | \Psi ' \rangle \propto \Lambda^3
\end{equation}
where $\Lambda$ is a scale related to the binding of the heavy
quarkonia. Here we may consider two extreme cases. The first one
is the purely coulombic scenario, in which the parameter
$\Lambda$ is the heavy mass times the fine structure
constant
\begin{equation}
\Lambda = \alpha_s (m_Q) m_Q
\end{equation}
and therefore one would expect a strong scaling with the heavy
mass.
This is, however, not supported by the energy differences in the
known
quarkonia, where e.g. the level spacing $M_{2S} - M_{1S}$
is practically the same in the $J/\psi$ system and the $\Upsilon$
system.
So we shall argue that a scenario in
which $\Lambda$ is independent of the mass of the heavy quark is
realistic, and consequently we would expect that $g$ and $g_1$
should scale in the same way as the parameters $a_0$ and $b_0$ in
the
quark level Lagrangian ${\cal L}_{quark}$. Since
$g_{b\bar{b}} / g_{c\bar{c}}$ is practically unity, we may assume
that
both couplings are of order one in the heavy mass expansion. This
in turn would lead to the conclusion that $g_1 / g$ should not
scale
as a power of the heavy mass. However, the coupling $g_1$ is larger
in the
$J/\psi$ system by a factor between two or three compared to the
$\Upsilon $ system. The only consistent interpretation of this
factor
is that
\begin{equation}
\frac{g_{1 \, b\bar{b}}}{g_{1 \, c\bar{c}}} =
\left[ \frac{\alpha_s (m_b)}{ \alpha_s (m_c)}\right]^{\kappa},
\hspace{2cm} \kappa \simeq 2 - 3
\end{equation}
Consequently the second term of our effective Lagrangian must be
governed
at the level of QCD by an operator with a sufficiently large
anomalous
dimension.

Let us finally also consider the chiral symmetry breaking
contribution
relevant for the decays $\psi(2S) \to J/\psi \pi$ and $\Upsilon(2S)
\to
\Upsilon (1S)\pi$. Interpreting this transition again in terms of a
dimension-3 operator for the heavy quarks one writes
\begin{equation}
{\cal L}^\prime_{quark} = a^\prime_0 (\bar{Q}_v \gamma_5 Q_v)
\mbox{Tr } \left[
           (U - U^\dagger) \right]
\end{equation}
The leading term in the heavy mass expansion will vanish for this
operator, since
$$
\bar{Q}_v \gamma_5 Q_v = {\cal O} (1/m_Q)
$$
and hence one expects the coefficients $g^\prime$ to scale as
\begin{equation}
\frac{g'_{b \bar{b}}}{g'_{c \bar{c}}} = \frac{m_c}{m_b}\simeq 0.3
\end{equation}
assuming again that the mass dependence of the matrix elements is
weak.

\section{Conclusions}
We have formulated an effective theory approach for the hadronic
decays
of heavy quarkonia based on the chiral limit for the light degrees
of
freedom and on the $1/m_Q$ expansion for the heavy ones. The
leading
term in both the inverse of the heavy quark mass and the inverse of
the chiral symmetry breaking scale consists of two terms multiplied
by coupling constants which we have fitted from experiment. The fit
to the invariant mass spectrum of the two pions in the decays
$\psi(2S) \rightarrow J/\psi\pi^+\pi^-$ and
$\Upsilon(2S)\rightarrow
\Upsilon(1S)\pi^+\pi^-$ shown in Fig.1 is quite satisfactory.

We also have discussed the mass dependence of the various terms in
the
Lagrangian which unfortunately may not be considered entirely in
the
effective theory. We model the leading terms of the effective
Lagrangian
by switching to a description in which the relevant degrees of
freedom
are pions and heavy quarks. This model resembles the chiral quark
model
as proposed by Georgi and Manohar \cite{ChiQua}. In this picture
the
coupling constants
of the quarkonia-pions effective Lagrangian may be reexpressed in
terms
of the ones of the underlying quark Lagrangian and certain matrix
elements
of heavy quark operators between heavy quarkonia states. The mass
dependence
of the matrix elements may not be evaluated from chiral and heavy
quark
symmetry and thus requires additional input. Using a specific
assumption
on the mass dependence of these matrix elements we arrive at
definite
predictions for the ratios  $g_{b\bar{b}}/g_{c\bar{c}}$ and
$g'_{b\bar{b}}/ g'_{c\bar{c}}$. The predictions are consistent with
what is found experimentally, although for $g'_{b\bar{b}}/
g'_{c\bar{c}}$
only an experimental bound exists and more data is needed to
confirm or
falsify our values.\\[8mm]
\section*{Acknowledgements}
R.U. thanks H.Leutwyler for discussions and H.Genz for
providing unpublished notes.

\end{document}